\documentclass[twoside,twocolumn,9pt]{article}
\usepackage{extsizes}
\usepackage[super,sort&compress,comma]{natbib} 
\usepackage[version=3]{mhchem}
\usepackage[left=1.5cm, right=1.5cm, top=1.785cm, bottom=2.0cm]{geometry}
\usepackage{balance}
\usepackage{mathptmx}
\usepackage{sectsty}
\usepackage{graphicx} 
\usepackage{lastpage}
\usepackage{amsmath}
\usepackage{upgreek}
\usepackage{natbib}
\usepackage{SIunits}
\usepackage[super]{nth}
\usepackage[format=plain,justification=justified,singlelinecheck=false,font={stretch=1.125,small,sf},labelfont=bf,labelsep=space]{caption}
\usepackage{float}
\usepackage{fancyhdr}
\usepackage{fnpos}
\usepackage[english]{babel}
\addto{\captionsenglish}{%
  
}
\usepackage{array}
\usepackage{droidsans}
\usepackage{charter}
\usepackage[T1]{fontenc}
\usepackage[usenames,dvipsnames]{xcolor}
\usepackage{setspace}
\usepackage[compact]{titlesec}
\usepackage{hyperref}

\usepackage{epstopdf}

\definecolor{cream}{RGB}{222,217,201}

\begin{document}

\pagestyle{fancy}
\thispagestyle{plain}
\fancypagestyle{plain}{
\renewcommand{\headrulewidth}{0pt}
}

\makeFNbottom
\makeatletter
\renewcommand\LARGE{\@setfontsize\LARGE{15pt}{17}}
\renewcommand\Large{\@setfontsize\Large{12pt}{14}}
\renewcommand\large{\@setfontsize\large{10pt}{12}}
\renewcommand\footnotesize{\@setfontsize\footnotesize{7pt}{10}}
\makeatother

\renewcommand{\thefootnote}{\fnsymbol{footnote}}
\renewcommand\footnoterule{\vspace*{1pt}%
\color{cream}\hrule width 3.5in height 0.4pt \color{black}\vspace*{5pt}} 
\setcounter{secnumdepth}{5}

\makeatletter 
\renewcommand\@biblabel[1]{#1}            
\renewcommand\@makefntext[1]%
{\noindent\makebox[0pt][r]{\@thefnmark\,}#1}
\makeatother 
\renewcommand{\figurename}{\small{Fig.}~}
\sectionfont{\sffamily\Large}
\subsectionfont{\normalsize}
\subsubsectionfont{\bf}
\setstretch{1.125} 
\setlength{\skip\footins}{0.8cm}
\setlength{\footnotesep}{0.25cm}
\setlength{\jot}{10pt}
\titlespacing*{\section}{0pt}{4pt}{4pt}
\titlespacing*{\subsection}{0pt}{15pt}{1pt}

\fancyfoot{}
\fancyfoot[LO,RE]{\vspace{-7.1pt}\includegraphics[height=9pt]{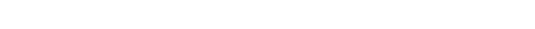}}
\fancyfoot[CO]{\vspace{-7.1pt}\hspace{13.2cm}\includegraphics{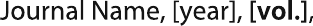}}
\fancyfoot[CE]{\vspace{-7.2pt}\hspace{-14.2cm}\includegraphics{RF}}
\fancyfoot[RO]{\footnotesize{\sffamily{1--\pageref{LastPage} ~\textbar  \hspace{2pt}\thepage}}}
\fancyfoot[LE]{\footnotesize{\sffamily{\thepage~\textbar\hspace{3.45cm} 1--\pageref{LastPage}}}}
\fancyhead{}
\renewcommand{\headrulewidth}{0pt} 
\renewcommand{\footrulewidth}{0pt}
\setlength{\arrayrulewidth}{1pt}
\setlength{\columnsep}{6.5mm}
\setlength\bibsep{1pt}

\makeatletter 
\newlength{\figrulesep} 
\setlength{\figrulesep}{0.5\textfloatsep} 

\newcommand{\topfigrule}{\vspace*{-1pt}%
\noindent{\color{cream}\rule[-\figrulesep]{\columnwidth}{1.5pt}} }

\newcommand{\botfigrule}{\vspace*{-2pt}%
\noindent{\color{cream}\rule[\figrulesep]{\columnwidth}{1.5pt}} }

\newcommand{\dblfigrule}{\vspace*{-1pt}%
\noindent{\color{cream}\rule[-\figrulesep]{\textwidth}{1.5pt}} }

\makeatother

\twocolumn[
  \begin{@twocolumnfalse}
{\includegraphics[height=30pt]{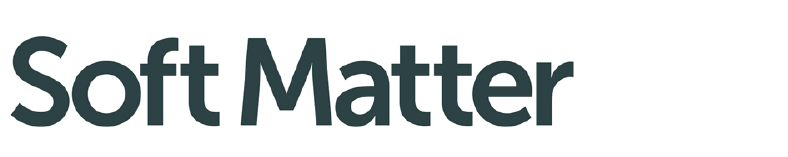}\hfill\raisebox{0pt}[0pt][0pt]{\includegraphics[height=55pt]{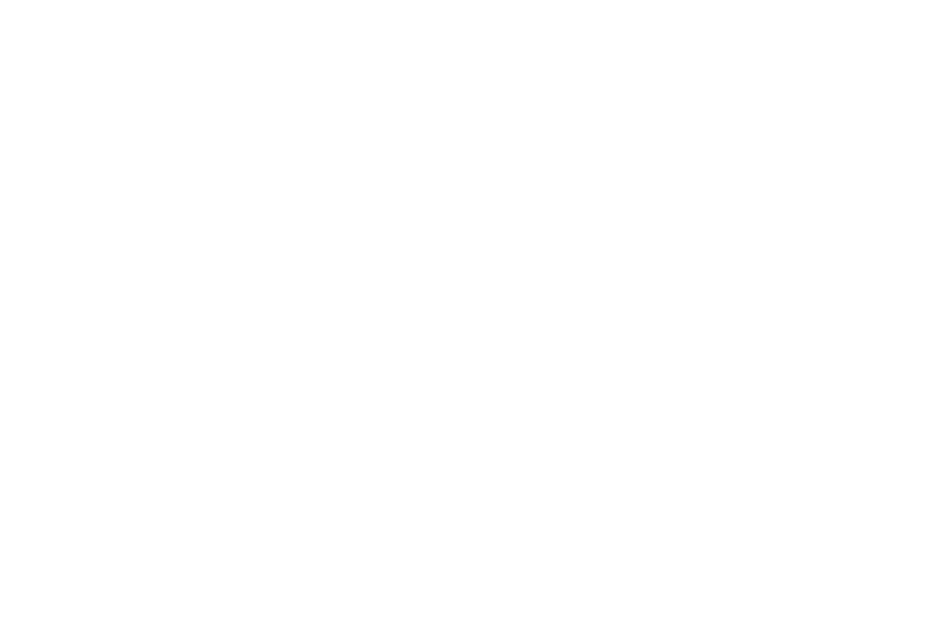}}\\[1ex]
\includegraphics[width=18.5cm]{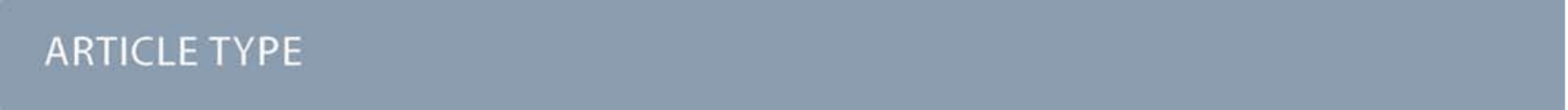}}\par
\vspace{1em}
\sffamily
\begin{tabular}{m{4.5cm} p{13.5cm} }

\includegraphics{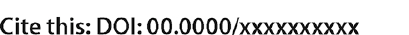} & \noindent\LARGE{\textbf{Nonlinear amplification of adhesion forces in interleaved books}} \\
\vspace{0.3cm} & \vspace{0.3cm} \\

 & \noindent\large{Raphaelle Taub,\textit{$^{a}$} Thomas Salez,\textit{$^{c,d}$}  Héctor Alarc\'{o}n,\textit{$^{b}$} Elie Rapha\"{e}l,\textit{$^{e}$} Christophe Poulard,$^{\ast}$\textit{$^{a}$} and Frédéric Restagno$^{\ast\ast}$\textit{$^{a}$}} \\

\includegraphics{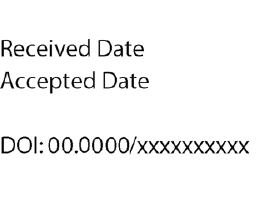} & \noindent\normalsize{ Two interleaved phonebooks are nearly impossible to separate by pulling their spines. The very slight force exerted by the outer sheets of the assembly is amplified as the exponential of the square of the number of sheets, leading to a highly-resistant system even with a small number of sheets. We present a systematic and detailed study of the influences of the normal external force and the geometrical parameters of the booklets on the assembly strength. We conclude that the paper-paper adhesion force between the two outer sheets, on the order of a few \milli\newton,is the one that is amplified by the two interleaved system. The two-phonebooks experiment that has attracted the attention of students and non-scientific public all around the world as an outstanding demonstration of the strength of friction appears to be as well a spectacular macroscopic manifestation of the microscopic friction-adhesion coupling.} \\

\end{tabular}

 \end{@twocolumnfalse} \vspace{0.6cm}

  ]

\renewcommand*\rmdefault{bch}\normalfont\upshape
\rmfamily
\section*{}
\vspace{-1cm}


\footnotetext{\textit{$^{a}$}~Université Paris-Saclay, CNRS, Laboratoire de physique des solides, Orsay, France.}
\footnotetext{\textit{$^{b}$}~Instituto de Ciencias de la Ingener\'{i}a, Universidad de O'Higgins, Rancagua, Chile.}
\footnotetext{\textit{$^{c}$}~Univ. Bordeaux, CNRS, LOMA, UMR 5798, F-33405 Talence, France.}
\footnotetext{\textit{$^{d}$}~Global Station for Soft Matter, Global Institution for Collaborative Research and Education,
Hokkaido University, Sapporo, Hokkaido 060-0808, Japan.}
\footnotetext{\textit{$^{e}$}~UMR CNRS 7083 Gulliver, ESPCI Paris, PSL Research University, Paris, France.}
\footnotetext{$^{\ast}$~Corresponding author. Email: christophe.poulard@universite-paris-saclay.fr}
\footnotetext{$^{\ast\ast}$~Email: frederic.restagno@universite-paris-saclay.fr}





Solid friction is a classical problem both in daily life and engineering but is also a fundamental question \cite{persson_sliding_2000}. In their  pioneering works, Coulomb and Amontons developed what is now called the Coulomb-Amontons laws of friction. They introduced a friction coefficient which is the ratio between the traction force required to be at the onset of sliding and the normal force between two solids. It took a long time to understand the molecular mechanisms of friction. Tabor was the first to identify how the adhesive junction between the microscopic surface's asperities could be responsible of one of the most non-trivial characters of the friction coefficient: it is independent of the apparent contact area between the two sliding surfaces \cite{bowden_friction_2001}. More recently, a lot of efforts have been put to understand friction at the micro and nanoscales \cite{bhushan_nanotribology:_1995,berman_macroscale_2015,egberts_frictional_2014}, in biology \cite{ward_solid_2015,ghosal2012capstan} and meta-materials \cite{ma_anisotropic_2019}.
Among the open questions of tribology remains the understanding of the link between adhesion and friction \cite{yoshizawa_fundamental_1993,israelachvili_relationship_1994,chateauminois_local_2008}. Biological systems such as gecko's toes show a  precise coupling of the normal and tangential forces due to adhesion that is important in particular to understand the detachment mechanisms \cite{tian_adhesion_2006,yu_friction_2012,stark_surface_2013,zhao_adhesion_2008}.
In engineering, another example of the importance of tuning adhesion in order to control friction is the haptic device, where an electric field allows to modify the adhesion between fingers and a touch screen in order to control their friction, generating a sensible mechanical stimulation \cite{hayward_new_2000,richard_friction_2002}. 
Systems exhibiting a lot of frictional contacts could be used in order to understand better how adhesion could modify global macroscopic performances. From a physical perspective, granular materials where a small amount of humidity can strongly affect the mechanical properties \cite{herminghaus_wet_2013,raux_cohesion_2018,restagno_aging_2002} are important, and in daily life, braids \cite{hristov_mechanical_2004}, knitted fabrics \cite{warren2018clothes,poincloux2018geometry}, and interlocked chains or fibers \cite{dumont2018emergent} are widespread.

Another example of these common, yet puzzling, systems is the popular demonstration of the strength of friction using two  phonebooks that are interleaved page-by-page and pulled apart by their spines \cite{dalnoki-veress_why_2016}. A simple model that captures the force necessary to separate the books as a function of their interleaving distance and the number of sheets has been presented in \cite{alarcon2016self} and experimentally tested on controlled paper assemblies. The main idea of the model is that the minute friction force $T^{*}$ exerted by the outer sheet of the assembly on the sheets below, is amplified because of the inherent angle present in the interleaved geometry, that induced a conversion of the operator traction into a supplementary normal force -- and thus additional friction. This is a mechanism similar to the well-known amplification of the tension force created with a capstan. This led to link the pulling force $T$ exerted on the whole assembly, to the separation distance $d$ between the booklets  and the number $2M$ of sheets by booklet, as \cite{alarcon2016self}:  
{\begin{equation}
T=2MT^{*}\sqrt{\frac{\pi}{4\alpha}}\text{exp}(\alpha)\text{erf}(\sqrt{\alpha}),     
\label{eq} \end{equation} }
where $\alpha$ is the \textit{Hercules number} defined as $\alpha=2\mu \epsilon M^2/d$, $\mu$ is the friction coefficient and $\epsilon$ is the thickness of one sheet.

\begin{figure}[t!]
\centering     
\includegraphics[width=9cm]{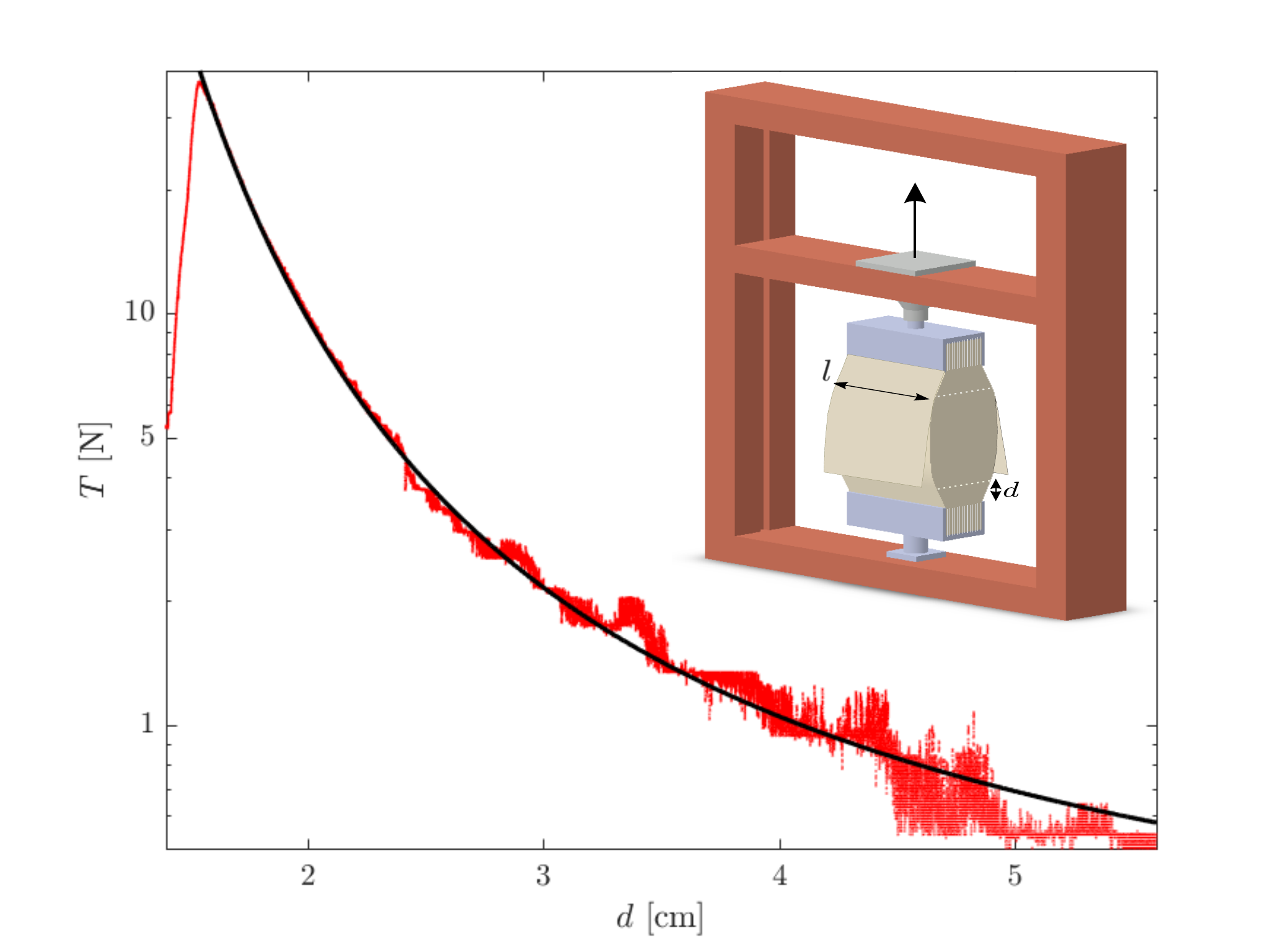}     
\caption{\textbf{Traction force exerted on the two-booklets assembly as a function of interleaved distance,} measured for two booklets of $2M=30$ sheets each, which dimensions are $L$=\unit{16}{\centi\meter}, $l$=\unit{21}{\centi\meter} and $\epsilon$=\unit{0.1}{\milli\meter}. The solid line represents the best fit to ~\ref{eq}. The calculated fitting parameters are $\mu=2.4$ and $T^{*}$=\unit{4}{\milli\newton}.The traction force $T$ is recorded by a traction-force machine in which the booklets are clamped (see inset).}
\label{master} 
\end{figure}
Alarcon \emph{et al.}~treated the parameters $\mu$ and $T^{*}$ as fitting parameters \cite{alarcon2016self}. However, $T^{*}$ was a parameter with a direct impact on the mechanical properties of the assembly. 
The proportionality relation between $T^{*}$ and $T$ predicted by Eq.~\ref{eq} remains to be validated and the origin of $T^{*}$ has to be investigated. Moreover, the physical origin of $T^{*}$ should be elucidated.  Several forces could contribute significantly to $T^{*}$: the weight of the outer sheet or anything that is attached to it; the bending elastic force, due to the angle of the interleaved assembly; or the intermolecular interaction between the last sheet and the sheet below, which is the essence of adhesion.

In order to elucidate  further the role and origin of $T^{*}$, a systematic experimental study is carried out by carefully interleaving sheet by sheet two paper stacks. The assembly is blanketed on each side by a rigid cover in order to avoid any  important bending of the external sheet. Both booklets and the two covers are clamped in metallic jaws and fixed vertically into a traction-force machine (Adamel Lhomargy DY32). The total traction force $T$ is then measured (with an accuracy of \unit{0.1}{\newton}), as a function of the separation distance $d$ (measured with an accuracy of \unit{10}{\micro\meter}) between the clamp and the contact area (see Fig. \ref{master}). The two booklets are pulled apart at a constant and tunable speed varying from \unit{1}{\milli\meter\per\min} to \unit{10}{\milli\meter\per\min}.
Each booklet is made of the same number of identical sheets of paper (Inacopia Office\textsuperscript{TM}, \unit{80}{\gram\per\square\meter}, "silky touch"), with length $L$, and width $l$ that are can be varied for the experiments. The thickness of a sheet is $\epsilon=$\unit{0.1}{\milli\meter} and kept constant through all experiments. 
A special attention is given to the initial separation distance $d_0$ of the booklets, which is measured after clamping them into the traction force machine. For a given set of experiments, $d_0$ is maintained as constant as possible, since the experiments were found to be less reproducible for initial separation distances that varied over more than \unit{5}{\milli\meter}.

\begin{figure}[t!]
\centering     
\includegraphics[width=8cm]{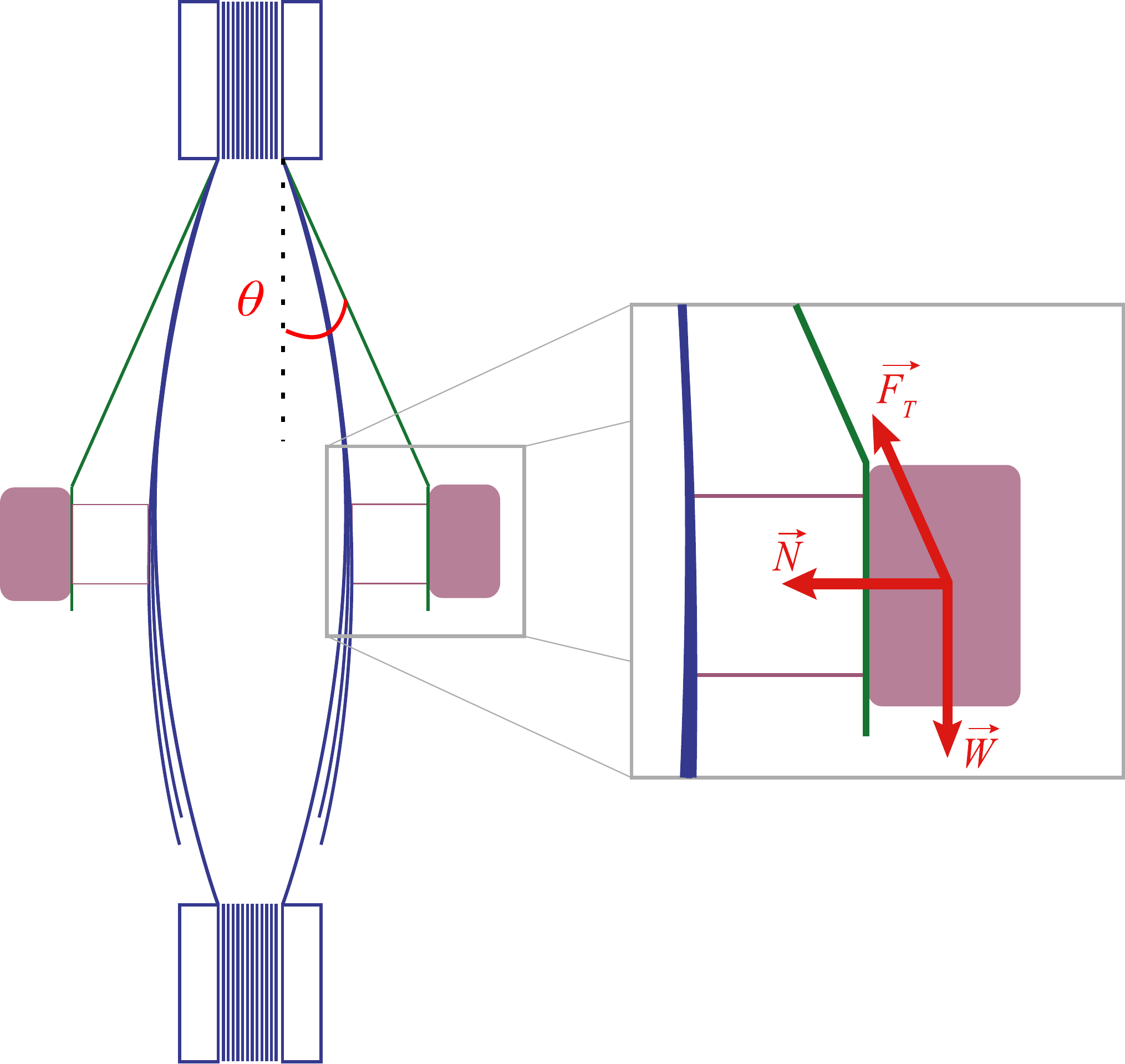}     
\caption{\textbf{Side view of the two-booklet assembly set with a tunable external load.} The assembly is covered on each side by two additional sheets (see inset): the first one (in green) is rigid cardboard and the second one (in blue)  composed of soft PTFE. Two masses weighing $m$ grams are fixed to the first cover, which is attached to the upper clamp. A spacer is inserted between the masses and the soft cover, and fixed to the rigid cover, in order to maximize the force. }
\label{profil_masse} 
\end{figure}

Using the previously described setup, a first  experiment is designed to drastically change the boundary condition by adding an external load on the cover. A spacer and a second rigid cover holding an adjustable load are added atop the first cover, on each side of the assembly (see Fig.~\ref{profil_masse}). A soft polytetrafluoroethylene (PTFE) sheet is further added between the spacer and the first cover, in order to minimize the stick-slip of the external load. The spacer allowed to maximize the normal load exerted on the assembly, by increasing the angle $\theta$ and thus the horizontal component of the weight-induced tension in the second cover.

\begin{figure}[t!]   
\centering     
\includegraphics[width=8cm]{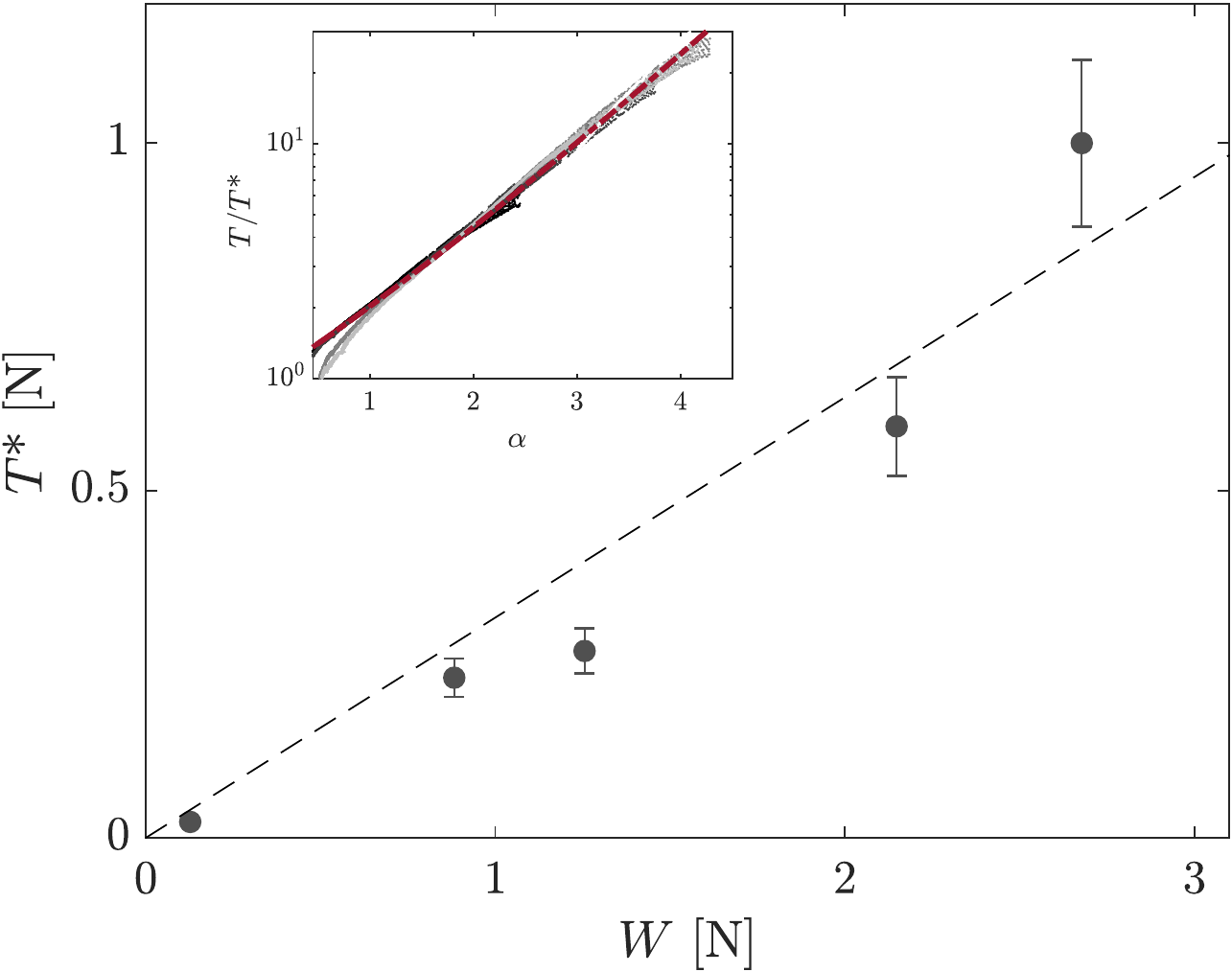}     
\caption{\textbf{Boundary friction force $T^{*}$ as a function of external weight.} 
The dashed line corresponds to the best linear fit with a slope of $0.32\pm0.1$.  The error on $T^{*}$ that is computed by roducing the experiment is 11~\%, and this value is used to generate error bars on the figure. The external weight is the weight of the entire fixing device with the tunable load, plus the weight of the second cover. The inset shows $T$/$T^{*}$ as a function of $\alpha$ for the different tested masses, as well as a best fit (dashed red line) to Eq.~\ref{eq} with $\mu$ and $T^{*}$ as free parameters.}
\label{t_m} 
\end{figure}

Force-displacement curves are recorded using this design, with various external masses $m$. Those $T(d)$ curves are well fitted by Eq.\ref{eq} with $\mu$ and $T^{*}$ as free parameters. The measured values of $T^{*}$ allow to rescale all the curves as plotted in the inset of Fig. \ref{t_m}, self-consistently confirming that the model developed without external loading is still valid.
Furthermore, as shown in Fig. \ref{t_m}, $T^{*}$ is found to depend on the external mass and thus on the effective external load. Indeed, when adding a small weight, $T^{*}$ clearly changes by more than two orders of magnitude. More precisely, $T^*$ varies linearly with the applied load. The best linear fit for the overall data is shown as a dashed line in Fig. \ref{t_m} and writes $T^*=(0.32\pm0.01)W$.
To understand further this dependency, one can consider the force balance on the outer mass at rest: $\overrightarrow{F_{T}}-\overrightarrow{N}+\overrightarrow{W}=\overrightarrow{0}$, where $\overrightarrow{F_{T}}$ is the tension force in the second cover on which the mass is fixed, $\overrightarrow{N}$ is the normal force exerted by the mass on the booklet assembly and $\overrightarrow{W}$ is the weight (see Fig. \ref{profil_masse}).
By taking into account the angle $\theta$ between the rigid cover and the vertical axe, a combination of the vertical and horizontal projections gives $N=W\tan{\theta}$. As $T^{*}$ is the friction force on the outer sheet, we have $T^{*}=\mu N$ from the Amontons-Coulomb law at the onset of the motion, which leads to $T^{*}=\mu W\tan{\theta}$. The measured $\theta$ is around $15$$^{\circ}$. Taking $\mu=1.07$, which is the mean $\mu$ over the experiments of Fig. \ref{t_m}, one gets $T^{*}=(0.29\pm0.1) W$. This is in good agreement with the experimentally  measured slope in Fig. \ref{t_m}. Therefore, it seems reasonable to conclude that in this case $T^{*}$ originates from and is proportional to the external mass. More generally, our results extend the validity of Eq.~\ref{eq} and the conclusion of Alarcon \textit{et al.} stating that the total traction force $T$ results from a frictional and geometrical amplification of $T^*$ \cite{alarcon2016self}.

\begin{figure}[t!]    
\centering     
\includegraphics[width=8cm]{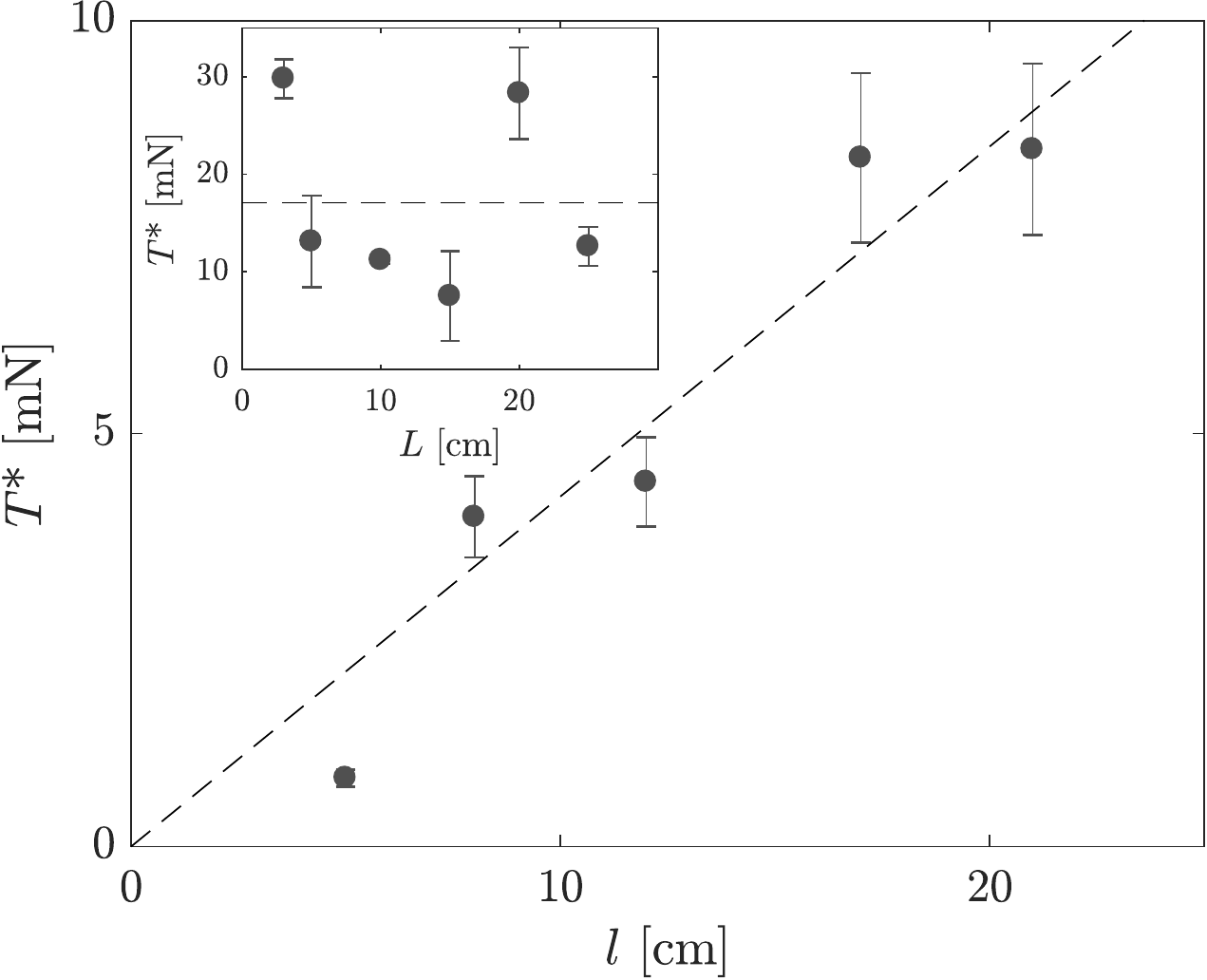}     
\caption{\textbf{Boundary normal force $N=T^{*}/\mu$ as a function of sheet width.} Using booklets with a constant length $L$=\unit{16}{\centi\meter} and various widths $l$, force-displacement curves are recorded and fit to Eq.~\ref{eq}, with $T^{*}$ and $\mu$ as free parameters. The best linear fit represented as a dashed line has a slope of \unit{42}{\milli\newton\per\meter}. The inset shows the values of $T^{*}$ for booklets with the same width $l$=\unit{12}{\centi\meter} and different lengths $L$.}
\label{tmul} 
\end{figure}

In order to determine the origin of the boundary force $T^{*}$ when no external mass is added, several experiments are performed with booklets made using sheets of different dimensions but keeping the total number of sheets in the assembly at $4M=60$.
First, the sheet width $l$ is varied from \unit{5}{\centi\meter} to \unit{21}{\centi\meter} while the length $L$ remains constant and equal to \unit{16}{\centi\meter} (see Fig.\ref{tmul}).  We observe that $T^{*}$ varies linearly with $l$. Secondly, the sheet length $L$ is varied from $3$ to \unit{25}{\centi\meter} with the width $l$ being kept at $l=\unit{12}{\centi\meter}$ (see inset of Fig. \ref{tmul}). In this case, $T^{*}$ is found to be scattered but essentially independent of $L$. \\
A first possible origin for $T^*$ could be the mass of the first cover.  This hypothesis is clearly wrong since it would imply the proportionality of $T^{*}$ with both $L$ and $l$. The order of magnitude is also not consistent:  $T$*$/l=\mu W\tan{\theta}/l$ is around \unit{1}{\milli\newton\per\meter}, which is ~40 times smaller than the measured value (see Fig.~\ref{tmul}).

A second possible origin of  $T^*$ is the elastic bending force. Indeed, the first cover is curved, and thus exerts a restoring elastic force on the assembly. The force required to bend a sheet depends on the boundary conditions clamping but scales as ~\cite{landau_theory_1999}: $E\epsilon^3lM\epsilon/x_0^3$ with $E$ the Young modulus of paper and $x_0$ the distance between the clamping point and the bending force application point. This force is proportional to the width of the page, like the measured $T^*$. Nevertheless, the distance $x_0$ is not obvious. It must be in between $d$ and $L$. The latter can be rejected due to the fact that we do not observe any dependency of $T^*$ with respect to $L$. We cannot have either $x_0=d$, despite the fact that it is the most intuitive hypothesis, since it would mean that the bending force changes during an experiment and thus we would not be able to fit our results with a single $T^*$ value. Similarly, any intermediate value between $d$ and $L$ is expected to depend on $d$, and should thus be rejected as well.

The last possible explanation is to take into account the adhesion energy between the two outermost sheets. Indeed, it is well known that the adhesive peeling force between two surfaces is proportional to the contact width $l$ and the energy release rate $G$ \cite{kaelble_theory_1959,dalbe_multiscale_2015,poulard_mechanical_2011}. Since we are interested in a quasi-static situation, 
the dissipation processes can be neglected, and the energy release rate $G$ is identified to the work of adhesion $w_{\textrm{A}}$. For a paper-paper symmetrical interfacial rupture, we have $w_{\textrm{A}}=2\gamma_{\textrm{p}}$ where $\gamma_{\textrm{p}}$ is the interfacial tension. Considering rough contacts, we can estimate the effective adhesion energy as $w_{\textrm{A,eff}}$ $=2\gamma_{\textrm{p}} A/A_0$ where $A/A_0$ is the fraction of the apparent area $A_0$ effectively involved in the contact. Invoking the interfacial tension $\gamma_{\textrm{p}}\sim29.6$~mN/m of hydrophobic cellulose fibers \cite{forsstrom_new_2005}, with the mean value of $\mu$=2.5 as a fitting parameter for the experiments at low masses in Fig.\ref{tmul}, the effective adhesion force per unit of width is thus comparable to the value of $N/l=T^{*}/(\mu l)=$\unit{16.8}{\milli\newton\per\meter} obtained in Fig.~\ref{tmul} if we assume $A/A_0\approx 0.28$, which is a plausible value for rough surfaces \cite{majumdar1991fractal}.

In conclusion, by investigating in a systematic manner the impacts of an external mass and the geometrical parameters of the booklets on the tearing force, and specifically on the boundary force $T^{*}$, we were able to identify the origin of the latter. The linear dependency of $T^{*}$ with the external mass confirmed further that Eq.~\ref{eq} and the underlying model based on a frictional and geometrical amplification of the boundary force captures well the phenomenon. This further suggests that the interleaved booklets assembly could be finely controlled in its resistance by an external load, and thus not only used as a toy experiment for physics demonstrations but also as a mechanical transistor, the outer mass acting as the gate terminal. This type of device has recently attracted a lot of attention from the soft robotic community \cite{rafsanjani2018kirigami,miriyev2017soft}.\\
We also evidenced a linear increase of $T^{*}$ with the booklets's width. Combined with the independence of $T^{*}$ in the books' length, this observation allowed us to show  that, in the absence of an external load, the main contribution to $T^{*}$ could neither be the weight of the cover nor the bending of the cover sheet. Finally, we suggested that the physical origin of the force exerted by the cover is its adhesion to the sheet. 
Further experiments using the same system of interleaved sheets, with different materials involving lower or higher adhesion forces would provide an interesting way of tuning the strength of the assembly.

\section*{Acknowledgements}
It is our great pleasure to thank K. Dalnoki-Veress for continuous discussions about this problem. We thank C. Aujoux for unpublished preliminary results, and Mason Porter for suggesting the name ``Hercules number" via the Improbable Research blog. We benefited from the financial support of the ANR (ANR-17-CE08-0008).



\balance


\bibliography{Abiblio} 
\bibliographystyle{rsc} 

\end{document}